\documentclass[a4paper,12pt]{article}
\usepackage[pctex32]{graphics}

\begin{document}
\topmargin 0pt \oddsidemargin 0mm

\renewcommand{\thefootnote}{\fnsymbol{footnote}}
\begin{titlepage}
\begin{flushright}
INJE-TP-07-01\\
gr-qc/yymmnnn
\end{flushright}

\vspace{5mm}
\begin{center}
{\Large \bf Black hole and  holographic dark energy} \vspace{12mm}

{\large   Yun Soo Myung\footnote{e-mail
 address: ysmyung@inje.ac.kr}}
 \\
\vspace{10mm} {\em  Institute of Mathematical Science and School
of Computer Aided Science \\ Inje University, Gimhae 621-749,
Korea }

\end{center}

\vspace{5mm} \centerline{{\bf{Abstract}}}
 \vspace{5mm}
We discuss the connection between black hole and holographic dark
energy.  We examine the issue of the equation of state (EOS) for
holographic energy density as a candidate for the dark energy
carefully.  This is closely related to   the EOS for black hole,
because the holographic dark energy comes from the black hole
energy density. In order to derive the EOS of a black hole, we may
use its dual (quantum) systems. Finally, a regular black hole
without the singularity is introduced to describe an  accelerating
universe inside the cosmological horizon. Inspired by this, we
show that the holographic energy density with the cosmological
horizon as the IR cutoff leads to the dark energy-dominated
universe with $\omega_{\rm \Lambda}=-1$.

\end{titlepage}
\newpage
\renewcommand{\thefootnote}{\arabic{footnote}}
\setcounter{footnote}{0} \setcounter{page}{2}

\section{Introduction}
Observations of supernova type Ia suggest that our universe is
accelerating~\cite{SN}. Considering the ${\rm \Lambda}$CDM
model~\cite{SDSS,Wmap1}, the dark energy and cold dark matter
contribute $\Omega^{\rm ob}_{\rm \Lambda}\simeq 0.74$ and
$\Omega^{\rm ob}_{\rm CDM}\simeq 0.22$ to the critical density of
the present universe. Recently the combination of WMAP3 and
Supernova Legacy Survey data shows a significant constraint on the
EOS for the dark energy, $w_{\rm ob}=-0.97^{+0.07}_{-0.09}$ in a
flat universe~\footnote{Another combination of data shows $w_{\rm
ob}=-1.04 \pm 0.06$~\cite{SSM}.}~\cite{WMAP3}.

Although there exist a number of dark energy models~\cite{CST},
the two promising candidates are the cosmological constant  and
the quintessence scenario. The EOS for the latter is determined
dynamically by the scalar or tachyon. In the study of dark
energy~\cite{UIS}, the first issue is whether the dark energy is a
cosmological constant with $\omega_{\Lambda}=-1$. If the dark
energy is shown not to be a cosmological constant, the next  is
whether the phantom-like state of $\omega_{\Lambda}<-1$ is
allowed. However, most theoretical models that may lead to
$\omega_{\Lambda}<-1$ confront with serious problems including
violation of the  causality. The last issue is whether
$\omega_{\Lambda}$ is changing (dynamical) as the universe
evolves.

On the other hand, there exists another model of the dark energy
arisen from the holographic principle. The authors in~\cite{CKN}
showed that in quantum field theory, the ultraviolet (UV) cutoff
$\Lambda$ could be related to the infrared (IR) cutoff $L$ due to
the limit set by forming a black hole. If $\rho_{\rm
\Lambda}=\Lambda^4$ is the vacuum energy density caused by the UV
cutoff, the total energy for a system of size $L$ should not
exceed the mass of the system-size black hole:
\begin{equation} \label{1eq1}
E_{\rm \Lambda} \le E_{BH}  \longrightarrow L^3 \rho_{\rm
\Lambda}\le  M_{\rm p}^2L.
\end{equation}
If the largest cutoff $L$ is chosen to be the one saturating this
inequality,  the holographic energy density is given by the energy
density of a system-size black hole as
\begin{equation} \label{1eq2}
\rho_{\rm \Lambda}=\frac{3c^2M_{\rm p}^2}{8\pi L^2} \simeq
\rho_{\rm BH},~~\rho_{\rm BH}=\frac{3M_{\rm p}^2}{8\pi L^2}
\end{equation}
 with a
constant $c$. Here we regard $\rho_{\rm \Lambda}$ as the dynamical
cosmological constant like the quintessence density of $\rho_{\rm
Q}=\dot{\phi}^2/2+V(\phi)$~\cite{UIS}. At the planck scale of
$L=M_{\rm p}^{-1}$, it is just the vacuum energy density
$\rho_{\rm V}=M^2_{\rm p}\Lambda_{\rm eff}/8 \pi$ of the universe
at $\Lambda_{\rm eff} \sim M^2_{\rm p}$: $\rho_{\rm \Lambda} \sim
\rho_{\rm p} \sim M^4_{\rm p}$. This implies that a very small
system has an upper limit on the energy density as expected in
quantum field theory. On the other hand, a larger system gets a
smaller energy density. If the IR cutoff is taken as the size of
the current universe ($L=H_0^{-1}$), the resulting energy density
is close to the current dark energy: $\rho_{\rm \Lambda} \sim
\rho_{\rm c}\sim 10^{-123}M^4_{\rm p}$~\cite{HMT}. This results
from the holography: the energy increases with the linear size, so
that the energy density decreases with the inverse-area law. The
total energy density dilutes as $L^{-3}$ due to the evolution of
the universe, whereas its upper limit set by gravity (black hole)
decreases as $L^{-2}$.
 Even though
it may explain the present data, this
 approach with $L=H_0^{-1}$ fails to recover the EOS for a dark
energy-dominated universe. This is  because there exists a missing
information about the pressure $p_{\rm \Lambda}$ of holographic
dark energy.

It is not easy to determine the EOS for  a system including
gravity with the UV and IR cutoffs. If one considers $L=H_0^{-1}$
together with the cold dark matter, the EOS may take the form of
$w_{\rm \Lambda}=0$~\cite{HSU}, which is just that of the cold
dark matter. However, introducing an interaction between
holographic dark energy and cold dark matter may lead to an
accelerating universe~\cite{Hor}.  Interestingly, the future event
horizon\footnote{As a concrete example, we introduce the
definition of the future event horizon $R_{\rm
FH}=a(t)\int_{t}^{\infty}\frac{dt'}{da(t')}$ with the
Friedmann-Robertson-Walker metric $ds^2_{\rm
FRW}=-dt^2-a^2(t)(d\tilde{r}^2+\tilde{r}^2d\tilde{\Omega}^2_2)$.
Assuming the power-law behavior of $a(t)=a_0
t^{\frac{2}{3(1+\omega_{\rm \Lambda})}}$~\cite{CTS}, one finds
$R_{\rm FH}=-\frac{3(1+\omega_{\rm \Lambda})}{1+3\omega_{\rm
\Lambda}}t$ for $-1<\omega_{\rm \Lambda}<-1/3$. In the case of
$a(t)=a_0 e^{Ht}$, one has $R_{\rm FH}=1/H$ with $\omega_{\rm
\Lambda}=-1$. This indicates that  de Sitter space can also
derived from the future event horizon.} was introduced to obtain
an accelerating universe~\cite{LI,HM,FEH,Myung2,KLM}.

At this stage, we emphasize  that the energy density $\rho_{\rm
BH}$ of the black hole is used to derive the holographic dark
energy. On the other hand, we do not use the pressure $p_{\rm BH}$
of the black hole to find the correct EOS of holographic dark
energy. Hence an important issue is to find the pressure of the
black hole.

In this Letter, we discuss a few of ways of obtaining  the EOS of
the black hole from its dual (quantum) systems. Further, we
introduce a regular black hole to obtain the dark energy from a
singularity-free black hole. Finally, we show that the holographic
energy density $\rho_{\rm \Lambda}$ with the cosmological horizon
leads to the dark energy-dominated universe with $\omega_{\rm
\Lambda}=-1$.

\section{EOS for black hole from dual (quantum) systems}
We start with the first law of thermodynamics
\begin{equation} \label{2eq1}
dE=TdS-pdV.
\end{equation}
On the other hand, the corresponding form of a non-rotating black
hole is given by
\begin{equation} \label{2eq2}
dE=TdS.
\end{equation}
The most conservative interpretation of $pdV=0$ is that the
pressure of a black hole vanishes, $p=0$.  This is consistent with
the integral form of $E=2TS$ (Euler relation).  If one chooses
$p_{\rm BH}=0$ really, the black hole plays a role of the cold
dark matter with
\begin{equation} \label{2eq3} w_{\rm BH}=0.
\end{equation}
It seems that the above  is consistent with the EOS $w_{\rm
\Lambda}=0$ for the holographic dark energy when choosing the
Hubble horizon $L=H_0^{-1}$~\cite{HSU}.

 As a non-zero pressure black hole, we may consider the AdS
black hole. In this case, we use the AdS-CFT correspondence to
realize the holographic principle~\cite{Witt}. In fact, we have
the dual holographic model of the boundary CFT without gravity.
Hence  we define the energy density and pressure on the boundary
by using the AdS-CFT correspondence. The EOS of CFT is given by
\begin{equation} \label{2eq4}
w_{\rm CFT}=\frac{1}{3}
\end{equation}
which shows that the CFT  looks like a radiation-like matter at
high temperature ~\cite{Verl}. It is suggested that the AdS black
hole may have the same EOS as that of CFT at high temperature.
This means that we could obtain the EOS of black hole
 at high temperature from its dual CFT through the AdS-CFT correspondence.

However, for the Schwarzschild black hole, the corresponding
holographic model is not yet found~\cite{KPSZ}. This may be so
because the Schwarzschild black hole is too simple to split the
energy into the black hole energy and Casimir energy, in contrast
to the AdS black hole~\cite{myungee}. Recently, there was a
progress on this direction. The authors~\cite{BS} showed that the
energy-entropy duality transforms a strongly interacting
gravitational system (Schwarzschild black hole) into a weakly
interacting quantum system (quantum gas). The duality
transformation between black hole ($E,~S,~T$) and dual quantum
system ($E^\prime,~S^\prime,~T^\prime$) is proposed as
\begin{equation} \label{2eq5} S^\prime \to E=M,
~E^\prime \to S=A/4,~T^\prime \to \frac{1}{T}=8 \pi M
\end{equation}
with $A=4 \pi M^2$. This may provide a hint for the
quantum-corrected  EOS of the Schwarzschild black hole. In this
case, they used an extensive thermodynamic relation
\begin{equation} \label{2eq6}
E=TS-pV
\end{equation}
which holds if the pressure is non-zero. A choice of the negative
pressure $p_{\rm QG}=-TS/V$ leads to
\begin{equation} \label{2eq7}
 E=2TS,
 \end{equation} which is just the case of the
black hole\footnote{This relation was proved to hold for a general
spherically symmetric horizon~\cite{Pad1}. Defining  the entropy
$S$ as a congruence (observer) dependent quantity and the energy
$E$ as the integral over the source of the gravitational
acceleration for the congruence, one recovers the relation
$S=E/2T$ between entropy, energy, and temperature. Also this
approach provides the quantum corrections to the
Bekenstein-Hawking entropy for all spherically symmetric
horizons.}. However, this pressure term does no enter into the
first law of Eq.(\ref{2eq2}). This is because they require a
constraint of $pdV=0$ to derive the underlying quantum model. As
the temperature is associated with the black hole thermodynamics,
the pressure of $p_{\rm QG}=-TS/V$  is related to the quantum
nature of the corresponding holographic model. Here we find  the
EOS for the quantum gas
\begin{equation} \label{2eq8}
w_{\rm QG}=-\frac{1}{2},
\end{equation}
which indicates a kind of the dark energy. If one chooses $w_{\rm
QG}$ as the EOS of the Schwarzschild black hole, this could
describe an accelerating universe of $w_{\rm QG}<-1/3$. However,
$\omega_{\rm QG}=-0.5$ is not close to the observation data
$\omega_{\rm ob}=-0.97^{+0.07}_{-0.09}$.

\section{${\rm \Lambda}$ black hole}
We discuss another issue of the  singularity on the holographic
energy density~\cite{KLL}. The holographic dark energy  states
that the universe is filled with the maximal amount of dark energy
so that our universe has become a black hole. However, an
intuitive evidence that this argument may  be wrong is that there
is no definite evidence that we are approaching a black hole
singularity anytime soon. In deriving the holographic energy
density in Eq.(\ref{1eq2}), we did not take into account the
singularity inside the event horizon seriously.

In order to avoid the singularity,  one may  introduce a regular
black hole called the de Sitter-Schwarzschild (${\rm \Lambda}$)
black hole~\cite{Dym}. Using a self-gravitating droplet of
anisotropic fluid of mass density $\rho_{\rm m}=\rho_{\rm V}
e^{-r^3/r_{\rm CH}^2r_{\rm EH}}$ with $r_{\rm
CH}=\sqrt{3/\Lambda_{\rm eff}}=1/H$ and $r_{\rm EH}=2m/M^2_{\rm
p}$, the conservation of the energy-momentum tensor
$T^{\mu}~_{\nu}={\rm diag}[\rho_{\rm m}, -p_{\rm r},
-p_{\bot},-p_{\bot}]$ leads to
\begin{equation} \label{3eq2}
p_{\rm r}=-\rho_{\rm m},~~p_{\bot}=-\rho_{\rm m}-r
\frac{\partial_{r}\rho_{\rm m}}{2}
\end{equation}
with the radial pressure $p_{\rm r}$ and tangential pressure
$p_{\bot}$. The Arnowitt-Deser-Misner mass  is defined by $m=4 \pi
\int^{\infty}_0\rho_{\rm m}r^2dr$.  If $p_{\bot}=0$, one finds the
zero gravity surface where  the gravitational repulsion balances
the gravitational attraction. Here one finds the solution
\footnote{For $r \to 0$, one has the de Sitter metric $ds^2_{\rm
dS}=-(1-H^2r^2)d\tau^2+(1-H^2r^2)^{-1}dr^2+r^2d\Omega^2_2$ with
$T_{\mu\nu} \simeq \rho_{\rm m} g_{\mu\nu}(\rho_{\rm m}=\rho_{\rm
V}=M_{\rm p}^2\Lambda_{\rm eff}/8\pi$), while for $r\to \infty$,
one finds the Schwarzschild metric $ds^2_{\rm S}=-(1-r_{\rm
EH}/r)d\tau^2+(1-r_{\rm EH}/r)^{-1}dr^2+r^2d\Omega^2_2$ with
$T_{\mu\nu} \simeq 0$. Hence for $m>m_{\rm c}$, one has two
horizons: outer event horizon $r_{\rm EH}$ and inner cosmological
horizon $r_{\rm CH}$. Actually, the ${\rm \Lambda}$ black hole
looks like the Reissner-Nordstrom black hole with replacing the
singularity by de Sitter space. } that includes de Sitter space
near $r=0$ and asymptotically Schwarzschild spacetime at
$r=\infty$.
\begin{figure}[t!]
   \centering
   \includegraphics{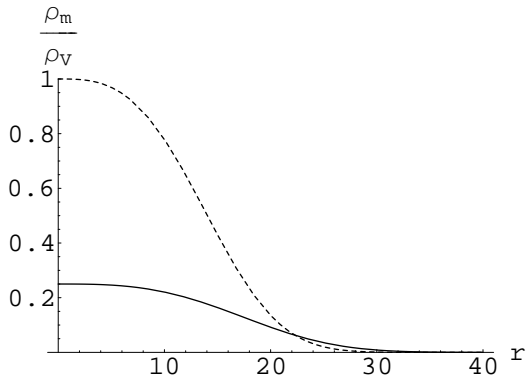}
   \caption{Plot of  density profile $\rho_{\rm m}/\rho_{\rm V}$ versus $r/(r_{\rm CH}^2r_{\rm EH})^{1/3}$
   with $\rho_{\rm V}=3/800\pi$ in the Planck units.
   The dashed curve is for the two horizons with $r_{\rm EH}=20$ and $r_{\rm CH}=10$,
   while the  solid curve is for the extremal black hole
   with $r_{\rm EH}=r_{\rm CH}=20$. Matter
   distribution is nearly flat both near the origin ($\rho_{\rm m} \simeq \rho_{\rm V}$) and for large
   $r(\rho_{\rm m} \simeq 0)$.} \label{fig}
   \end{figure}

As is shown in Fig. 1, the matter source $\rho_{\rm m}$ connects
smoothly de Sitter vacuum in the origin with the Minkowski vacuum
at infinity. For $m \ge m_{\rm c} \simeq 0.3M_{\rm
p}\sqrt{\rho_{\rm p}/\rho_{\rm V}}$, de Sitter-Schwarzschild
geometry describes a vacuum nonsingular black hole with $r_{\rm
EH}>r_{\rm CH}$, while for $m<m_{\rm c}$, it describes the G-lump
which is a vacuum self-gravitating  object without horizon. At
$m=m_{\rm c}$, we have the  extremal black hole with $r_{\rm
EH}=r_{\rm CH}$. Here de Sitter space replaces the singularity.
 In this case, we  have the EOS
\begin{equation} \label{3eq3}
w_{\rm dS}=-1,
\end{equation}
 inside the regular black hole.
Therefore we attempt to specify its EOS for the holographic dark
energy. If the radius of cosmological horizon $r_{\rm CH}$ is
taken to be the IR cutoff, one may consider the interior de Sitter
region to be a model of dark energy. Interestingly, the extremal
case  represents the limiting case when the Schwarzschild radius
of the system, whose size is the IR cutoff,  is equal to the IR
cutoff itself ($r_{\rm EH}=L=r_{\rm CH}$). However, two problems
 arise in this case. Any infinitesimal step towards a
non-saturated holographic dark energy would cause a sudden jump in
the EOS: from $-1$ to 0, so the EOS cannot be clearly determined.
Furthermore, the IR cutoff cannot be clearly determined because we
have the interior de Sitter space and thus, the Hubble distance
and the event horizon are degenerate. We note that the holographic
energy density $\rho_{\rm \Lambda}$ with $L=r_{\rm CH}$ is static
because $r_{\rm EH}$ is static. Thus, the holographic dark energy
approach is trivial for the $r_{\rm EH}=r_{\rm CH}$ case of
$\Lambda$ black hole.

In order to find a non-trivial case, we use the connection between
the static de Sitter space $(\tau,r)$ and  the dynamic
Friedmann-Robertson-Walker spacetime $(t,\tilde{r})$
\begin{equation} \label{3eq4}
\tau=t-\frac{1}{2H} \ln[1-H^2r^2],~~r=
\frac{\tilde{r}}{\sqrt{e^{2H\tau}+H^2 \tilde{r}^2}}.
\end{equation}
According to the Penrose diagram in Ref.\cite{FK}, their
asymptotic behaviors are closely related to each other. In  de
Sitter space, one has the future cosmological horizon $r_{\rm
CH}=1/H$ at $\tau=\infty$ only, while in the
Friedmann-Robertson-Walker space, one has the future event horizon
$R_{\rm FH}=1/H$ for $-\infty \le t \le \infty$.  In case of
$\tau=\infty$, a dynamical feature of $\rho_{\rm \Lambda}$ is
recovered and thus we have $\omega_{\rm \Lambda}=-1$. In this
sense,  the EOS of $\omega_{\rm dS}=-1$  is considered  to be  the
input and at most,  a consistency condition. Hence Eq.(\ref{3eq3})
is not considered as a derived result.

Inspired by this, we propose that the singular-free condition for
holographic dark energy $\rho_{\rm \Lambda}$ may determine the
equation of state.  As was pointed out at footnote 2, we  obtain
the de Sitter solution $L=r_{\rm CH}$ from the future event
horizon $R_{\rm FH}$. Here we choose the present universe-size
cosmological horizon as the IR cutoff~\cite{Pad2,MM}, which
contrasts to the case with the Hubble horizon
$L=1/H_0$~\cite{HSU}. For $L=1/H_0$, we could not determine its
EOS clearly, while for $ L=r_{\rm CH}=1/H$, we could determine its
EOS to be $w_{\rm \Lambda}=-1$.

\section{Discussions}
We are interested in the equation of state for black hole, because
the holographic dark energy   came from the energy density of
black hole. Here we wish to discuss the connection between the
black hole and holographic dark energy. Cohen et. al. [8]
mentioned that if one introduces the holographic principle, one
could include the gravity effects into the quantum field theory
naturally. This is because general relativity (black hole) is the
prime example of a holographic theory, whereas quantum field
theories are not holographic in their present form. The first
thing to realize holographic principle is given  by the
holographic  entropy bound which states that the entropy of the
system should be less or equal to the entropy of the system-size
black hole: $S_{\rm \Lambda}=L^3\Lambda^3 \le S_{\rm BH}=\pi
M_p^2L^2$. As was clarified by Cohen et.al., this bound includes
many states with $L_{\rm S}\sim L^{5/3}M_{\rm p}^{2/3} > L$.
Considering the energy $E_{\rm \Lambda}=L^3\Lambda^4$ of the
system  together with $\Lambda \sim (M_{\rm p}^2/L)^{1/3}$ (the
saturation of holographic  entropy bound), it implies $L_{\rm S}
\sim E_{\rm \Lambda}> L$ in the Planck units. This   shows a
contradiction that a larger black hole can be formed from the
system by gravitational collapse. Hence, one requires that no
state in the Hilbert space have energy so large that the
Schwarzschild radius $ L_{\rm S} \sim E_{\rm \Lambda}$ exceeds
$L$. Then,  a relation between the size $L$ of the system,
providing the IR cutoff and the UV cutoff $\Lambda$ is required to
be  Eq.(\ref{1eq1}) ($L_{\rm S} \sim E_{\rm \Lambda}< L$ in the
Planck units), which provides the constraint on $L$ that excludes
all states lying within $L_{\rm S}$. In physical terms, it
corresponds to the assumption that the effective field theory
describe all states of the system excluding those for which it has
already collapsed to a black hole. In other words, this relation
can be rewritten as $E_{\rm \Lambda} \le E_{\rm BH}$ called the
Bekenstein energy bound. This means that the energy of the system
should be less or equal to the energy of the system-size black
hole. Actually, both  holographic entropy bound and Bekenstein
energy bounds are based on the black hole.

If one takes the saturation of the energy bound in
Eq.(\ref{1eq2})(the limiting case) as the holographic dark energy
density, its EOS depends on the IR cutoff and/or interaction with
cold dark energy.

Let us calculate the average energy density $\rho$ of a
homogeneous spherical system that saturates the holographic
entropy bound. For this purpose,  we introduce the Bekenstein's
entropy bound $S \le 2 \pi E L$ which is another entropy bound. If
the system saturates the Bekenstein's entropy bound and
holographic entropy bound ($S= 2 \pi E L=S_{\rm BH}$), then it
satisfies the Schwarzschild condition of $E=M_{\rm p}^2L/2=E_{\rm
BH}$, which states that its maximal mass is the half of its radius
in the Planck units. The energy density $\rho$ is given by the
black hole energy density $\rho=E/V=3M_{\rm p}^2/8\pi
L^2=\rho_{BH}$, which is identical to the holographic energy
density $\rho_{\rm \Lambda}$ with $c^2=1$ shown in
Eq.(\ref{1eq2}). This shows the close connection between the black
hole and holographic dark energy.

 As was pointed out in~\cite{HM}, the
pressure of holographic dark energy  is determined  by the
conservation of energy-momentum tensor as $p_{\rm
\Lambda}=-\frac{1}{3}\frac{d\rho_{\rm \Lambda}}{d \ln a}-\rho_{\rm
\Lambda}$ which  provides the EOS of $\omega_{\rm
\Lambda}=\frac{p_{\rm \Lambda}}{\rho_{\rm
\Lambda}}=-1+\frac{2}{3H}\frac{\dot{L}_{\rm \Lambda}}{L}$.  Hence,
if one does not choose an appropriate form of $L$, one cannot find
its EOS. For example, if one chooses the Hubble horizon $L=1/H$,
it does not give the correct EOS~\cite{HSU}, but it leads to the
second Friedmann equation of
$\dot{H}=-\frac{3}{2}H^2(1+\omega_{\rm \Lambda})$. On the other
hand, choosing $L=R_{\rm PH/FH}$ leads to $\omega_{\rm
\Lambda}=-1/3(1\mp 2\sqrt{\Omega_{\rm \Lambda}}/c)$. Despite the
success of obtaining the EOS for $L=R_{\rm PH/FH}$, this may not
give us a promising solution to the dark energy problem because
choosing the future event horizon just means an accelerating
universe. That is, in order for the holographic dark energy to
explain the accelerating universe, we first must assume that the
universe is accelerating. This is not what we want to obtain
really: a realistic dark energy model will be determined from
cosmological dynamics with an appropriate EOS. In addition,
$\rho_{\rm \Lambda}$ violates causality because the current
expansion rate depends on the future expansion rate of the
universe. Thus  we may believe that taking the future event
horizon as the IR cutoff is just a trick to get an accelerating
universe in the holographic dark energy approach.

This attributes to the ignorance of the black hole pressure
because one uses mainly the energy density of the black hole to
describe the holographic dark energy. Hence we described how to
obtain the EOS of black holes from their dual systems as a first
step to understand the nature of holographic dark energy, although
it is still lacking for describing the pressure of the holographic
dark energy. In this approach, the limiting condition for the
saturated holographic energy density Eq.(\ref{1eq2}) is not found
for the EOS of the black hole from dual systems.

 Finally we consider the issue of the singularity
together with the holographic dark energy. In this direction, we
introduce the regular (${\rm \Lambda}$) black hole with two
horizons which includes de Sitter space near $r=0$ and
asymptotically Schwarzschild spacetime at $r=\infty$.  We find
that the singularity could be removed by choosing an appropriate
mass distribution and  de Sitter space appears inside the black
hole. However, we recover the dynamical behavior of holographic
energy density $\rho_{\rm \Lambda}$ with $L=1/H$ at $\tau=\infty$
because the static coordinates are used for calculation.

In conclusion, we  show that  the holographic dark energy without
the singularity lead to  the de Sitter-acceleration with
$\omega_{\rm \Lambda}=-1$.

\section*{Acknowledgment}
The author thanks Q. C. Huang for helpful discussions. This work
was in part supported by the Korea Research Foundation
(KRF-2006-311-C00249) funded by the Korea Government (MOEHRD) and
the SRC Program of the KOSEF through the Center for Quantum
Spacetime (CQUeST) of Sogang University with grant number
R11-2005-021.

\end{document}